\documentclass[12pt]{article}
\usepackage{graphicx} 
\usepackage{authblk}
\usepackage{amsmath}

\title{English translation of the paper ``Electromagnetic field inside a plane-parallel layer in the resonant absorption regime'' (1962) by A.~P.~Khapalyuk: Pioneering study on coherent perfect absorption}

\author{Denis V. Novitsky}
\affil{B.~I.~Stepanov Institute of Physics, National Academy of Sciences of Belarus, Nezavisimosti Avenue 68, 220072 Minsk, Belarus}

\date{\today}

\begin{document}
\maketitle

\begin{abstract}
In 1962, Belarusian physicist A.~P.~Khapalyuk has published the paper [Doklady Akademii nauk BSSR, volume 6, issue 5, pages 301-304] devoted to the effect of resonant absorption of light in a layer of matter. This work can be considered as a first study of the phenomenon known today as coherent perfect absorption (CPA). Unfortunately, the paper by A.~P.~Khapalyuk was published in a local journal in Russian and is almost unknown to the English-language scientific community. Here, we give the first translation of this paper into English, accompany it with the introductory remarks including some biographical information on A.~P.~Khapalyuk and put it in the context of modern CPA studies.
\end{abstract}

\section{Introductory remarks}

Coherent perfect absorption (CPA) is the effect of total absorption of coherent waves incident on an absorbing medium \cite{Baranov2017}. It is often treated as a time-reversed lasing, or anti-lasing, and is considered as an archetypal phenomenon of non-Hermitian physics. The idea of CPA is usually attributed to Y.~D.~Chong \textit{et al.} who published their seminal paper in 2010 \cite{Chong2010}. However, as noted by N.~N.~Rosanov \cite{Rosanov2017}, perhaps, the first study on CPA (under the name ``resonant absorption'') was published by the Soviet physicist A.~P.~Khapalyuk already in 1962. His paper appeared in the journal of Belarusian Academy of Sciences [Doklady Akademii nauk BSSR, volume 6, issue 5, pages 301-304] and was never translated into English. Attracted by this curious case of CPA  prehistory, we here address this gap in the literature and give the first English translation of that paper (see the second part of this manuscript). But first, we use this opportunity to give some biographical information of its author and to put his paper into the wider context of modern CPA studies.

\subsection{Biographical sketch of A.~P.~Khapalyuk}

Alexander Petrovich Khapalyuk was born on 13 March 1925 in Velikaryta village (then in Poland, now in Malaryta district, Brest region, Belarus). During the World War II, he fought the Germans in a partisan unit and from April 1944 in the ranks of the Red Army. He served in the army until 1950 with the rank of senior lieutenant \cite{poisk} and was awarded the medals ``For Courage'' and ``For the Victory over Germany in the Great Patriotic War 1941–1945'' \cite{VestnikBSU1}.

After demobilization, A.~P.~Khapalyuk entered the Faculty of Physics and Mathematics of the Belarusian State University in Minsk, where he became a student of Boris Ivanovich Stepanov -- the ``father'' of Belarusian optics and laser physics known for the international community mostly as the author of the universal relation between absorption and emission spectra of complex molecules (Kennard--Stepanov relation) \cite{Stepanov1957}. In 1955 A.~P.~Khapalyuk graduated from the university with honors and in 1959 completed his postgraduate studies. He worked as a junior researcher at the Institute of Physics of the Belarusian Academy of Sciences and taught at his alma mater: first as an assistant at the Department of Spectral Analysis and then as an Associate Professor at the Department of Physical Optics. In 1971, a new scientific division was established at the Belarusian State University -- the Institute for Applied Physical Problems. A.~P.~Khapalyuk was appointed the Head of Laboratory of Nonlinear Optics at this institute and spent there the rest of his career. In 1998, he resigned from his post and moved to the position of leading researcher at the Laboratory of Photonics \cite{VestnikBSU2}. In 1999, he has created and headed the unit of Belarusian Physical Society at his institute \cite{Kuchyna}.

Research of A.~P.~Khapalyuk is devoted to theoretical and mathematical aspects of optics and laser physics. In 1962 (the year he published his CPA paper), he defended the Ph.D. (Candidate of Sciences) thesis ``Optical properties of thin layer systems'', in which he developed the general theory of interference filters. Around the same time, the first lasers appeared facilitating a major change in the scientific interests of many optical physicists, including A.~P.~Khapalyuk. He has contributed widely to the theory of generation, propagation, and diffraction of laser beams. In particular, he studied the spatial structure of light field inside optical resonators filled with heterogeneous media. He published the monograph on this topic \cite{Belskii} and in 1987 defended the Doctor of Sciences thesis ``Open optical resonators and spatial structure of laser radiation''. He has also found the new solutions of Maxwell's equations within the framework of generalized functions and proposed the so-called kaleidoscopic spatial structures as a basis of new optical devices. He published almost 300 research papers (some of them can be found through Scopus \cite{Scopus}) and received more than 30 patents. 3 Doctor and 13 Candidate dissertations were defended under his supervision.

A.~P.~Khapalyuk died in 2010 -- the same year, when the modern history of CPA started.

\subsection{CPA then and now}

In his paper on resonant absorption, translation of which is given further, A.~P.~Khapalyuk considered the archetypal system of a single absorbing layer irradiated by two coherent waves from both sides. He obtained the conditions for total absorption due to interference of incident coherent waves. This is exactly the idea used in later works on CPA. However, the paper by Y.~D.~Chong \textit{et al.} \cite{Chong2010} and further studies contain a new important element -- the connection of this effect to the dynamics of scattering zeros and poles. This point of view proved to be fruitful to analyze different scenarios of light interaction with non-Hermitian systems and to identify a number of light scattering anomalies \cite{Krasnok2019}. In particular, CPA corresponds to a scattering zero appearing at the real frequency -- the situation provided by the presence of lossy medium in the system.

In fact, the idea of CPA can be traced even further and connected to the Borrmann effect known in the X-ray physics. In the 1940s, Gerhard Borrmann has found the anomalously weak absorption of X-rays in crystals due to mismatch between the positions of interference nodes and absorbing atoms \cite{Borrmann1941, Borrmann1950}. Optical analogs of the Borrmann effect were reported for periodic structures such as photonic crystals \cite{Novikov2017}. Moreover, CPA is sometimes directly associated with the Borrmann effect \cite{Liang2025}.

Thus, we have four different terms for the same phenomenon (CPA, anti-lasing, resonant absorption, and Borrmann effect) and can pose a question: which of the terms best describes it? N.~N.~Rosanov argues \cite{Rosanov2017} that ``resonant absorption'' is better than ``CPA'', especially since there is another meaning for CPA (chirped pulse amplification) widely used in the laser community. Although this opinion is well grounded, the common practice embodied in many reviews and research papers is unlikely to be changed. Anyway, the history of science is full of situations when perhaps not the best term became a standard one.

Not pretending to give a full picture of CPA studies and its width \cite{Baranov2017}, we want to end this brief discussion by mentioning several recent trends in this field. One of the trends is connected to studies of the structures containing both loss and gain, in particular PT-symmetric systems. The latter structures support both lasing and anti-lasing simultaneously \cite{Longhi2010, Wong2016} -- the so-called CPA-lasing effect. It is interesting how this effect is interrelated with the exceptional point (EP) degeneracies in such structures \cite{Novitsky2019} and can be observed even in non-PT-symmetric structures \cite{Novitsky2022}. Moreover, an EP where two or more eigenmodes coalesce can be combined with CPA when these two modes are incident at the lossy medium; this gives rise to the intriguing notion of CPA-EP \cite{Sweeney2019, Wang2021}.

Another trend is the employment of complex-frequency excitations resulting in the so-called virtual perfect absorption (VPA). Whereas CPA exploits a scattering zero by moving it at a real frequency with the help of a lossy medium, VPA can be observed in a lossless medium by using a scattering zero at a complex frequency. In a sense, the conditions for CPA and VPA can be considered as complementary: the former uses non-Hermitian (complex permittivity) media, whereas the latter deals with non-Hermitian (complex frequency) light. Since VPA implies lossless media, the complex-frequency radiation is not absorbed and leaves the medium after switching off excitation. The concept of VPA was first proposed in Ref. \cite{Baranov2017a}, generalized to elastodynamic \cite{Trainiti2019} and sound waves \cite{Maddi2025}, applied to resonant media \cite{Novitsky2023}, metasurfaces \cite{Marini2020}, and plasmas \cite{Delage2023} and is now considered as a representative of a wide class of complex-frequency effects \cite{Kim2025}.

\newpage

\section{Translation}

\begin{center}
{\Large \bf \textit{A.~P.~Khapalyuk}}

{\Large \bf Electromagnetic field inside a plane-parallel layer in the resonant absorption regime}

(\textit{Presented by academician B.~I.~Stepanov})
\end{center}

In Ref. (\footnote[1]{A.~P.~Khapalyuk, Inzhenerno-Phizicheskii Zhurnal (Journal of Engineering Physics and Thermophysics), issue 3, 1960.}), the solution of Maxwell's equations was obtained for the system of plane-parallel thin layers. Examination of the general formulas characterizing the optical properties of such systems shows that, when the coherent rays are simultaneously incident from both sides of the multilayer, there may be a case of the multilayer absorbing all incident light. To find out the conditions of such resonant absorption, let us consider the simplest case of a single plane-parallel thin slab.

\begin{figure}[t!]
\centering
\includegraphics[scale=0.8]{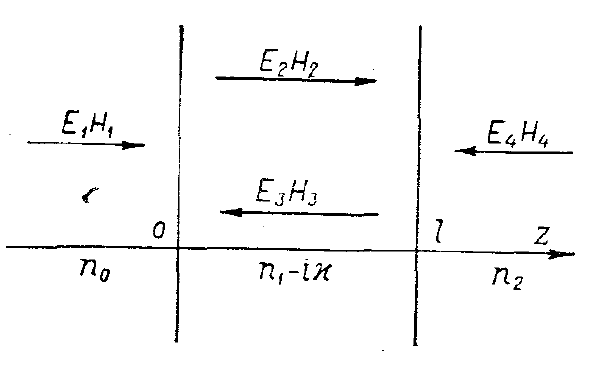}
\caption{\label{fig1}}
\end{figure}

To find the electromagnetic field at resonant absorption, it is enough to solve Maxwell's equations with the corresponding boundary conditions which take into account the absence of waves leaving the layer. Equations for the four plane waves propagating normally to the layer surface can be written as usual (Fig. \ref{fig1}):

\begin{eqnarray}
E_1=E_1^0 e^{i (\omega t - k n_0 z)}, \quad H_1=n_0 E_1^0 e^{i (\omega t - k n_0 z)}, \nonumber
\end{eqnarray}

\begin{eqnarray}
E_2=E_2^0 e^{-k \kappa z + i (\omega t - k n_1 z)}, \quad H_2=(n_1-i\kappa) E_2^0 e^{-k \kappa z + i (\omega t - k n_1 z)}, \label{eq1}
\end{eqnarray}

\begin{eqnarray}
E_3=E_3^0 e^{k \kappa z + i (\omega t + k n_1 z)}, \quad H_2=-(n_1-i\kappa) E_3^0 e^{k \kappa z + i (\omega t + k n_1 z)}, \nonumber 
\end{eqnarray}

\begin{eqnarray}
E_4=E_4^0 e^{i k n_2 l + i (\omega t + k n_2 z)}, \quad H_2=-n_2 E_4^0 e^{i k n_2 l + i (\omega t + k n_2 z)}, \nonumber
\end{eqnarray}

\noindent where $n_0$ and $n_2$ are the refractive indices of the media limiting the layer from both sides (these media are supposed to be transparent); $n_1$ and $\kappa$ are the real and imaginary parts of the complex refractive index; $k=2 \pi /\lambda$ is the wavenumber; $\lambda$ is the wavelength; $z$ is the coordinate normal to the layer plane and increasing in the direction from the $0$th medium to the $2$nd one; $l$ is the layer thickness.

Substituting (\ref{eq1}) into the boundary conditions requiring continuity of the tangential components of electric and magnetic vectors at both layer boundaries, we have

\begin{eqnarray}
(n_1-i\kappa) E_2^0 e^{-k \kappa l - i k n_1 l} - (n_1-i\kappa) E_3^0 e^{k \kappa l + i k n_1 l} + n_2 E_4 = 0, \nonumber
\end{eqnarray}

\begin{eqnarray}
E_2^0 e^{-k \kappa l - i k n_1 l} + E_3^0 e^{k \kappa l + i k n_1 l} - E_4 = 0, \nonumber
\end{eqnarray}

\begin{eqnarray}
n_0 E_1^0 - (n_1-i\kappa) E_2^0 + (n_1-i\kappa) E_3^0 = 0, \label{eq2}
\end{eqnarray}

\begin{eqnarray}
E_1^0 - E_2^0 - E_3^0 = 0. \nonumber
\end{eqnarray}

\noindent Nonzero solution of the system (\ref{eq2}) is possible only for the determinant equal to zero. From this, we obtain the conditions for resonant absorption

\begin{eqnarray}
(n_2 + n_1 - i\kappa) (n_1 + n_0 - i\kappa) e^{-k \kappa l - i k n_1 l} + \nonumber \\ + (n_2 - n_1 + i\kappa) (n_1 - n_0 - i\kappa) e^{k \kappa l + i k n_1 l} = 0. \label{eq3}
\end{eqnarray}

\noindent Complex equation (\ref{eq3}) is equivalent to the two real-valued ones

\begin{eqnarray}
e^{-4 k \kappa l} = r_1 r_2, \quad 2 k n_1 l = 2 \pi s - \rho_1 - \rho_2, \label{eq4}
\end{eqnarray}

\noindent where $r_1$ and $r_2$ are the Fresnel reflection coefficients,

\begin{eqnarray}
r_j = \frac{(n_j-n_1)^2+\kappa^2}{(n_j+n_1)^2+\kappa^2} \quad (j=1, 2); \label{eq5}
\end{eqnarray}

\noindent $\rho_1$ and $\rho_2$ are the phase shifts for the waves reflected at the layer boundaries,

\addtocounter{equation}{-1}
\begin{subequations}
\begin{eqnarray}
\rho_j = \arctan \frac{2 n_j \kappa}{n_j^2- n_1^2-\kappa^2} \quad (0 \leq \rho_j \leq \pi); \label{eq5a}
\end{eqnarray}
\end{subequations}

\noindent $s$ is any integer.

Conditions (\ref{eq4}) are the necessary requirements to the layer parameters -- the thickness $l$ and the refractive index ($n_1$, $\kappa$) making resonant absorption possible. Thus, the three layer parameters should satisfy the two equations (\ref{eq4}). It is natural that the layer parameters are ambiguous, so that one of them can be set arbitrarily within rather wide limits. For example, at the fixed layer thickness $l$, equations (\ref{eq4}) can be satisfied by choosing the real and imaginary parts of the layer refractive index, this choice being ambiguous even in this case (the parameter $s$ can have any value).

Excluding the layer thickness $l$ from equations (\ref{eq4}), we find the relation between the real and imaginary parts of the layer refractive index,

\begin{eqnarray}
\ln r_1 r_2 = 2 \frac{\kappa}{n_1} (2 \pi s' + \rho_1 + \rho_2), \label{eq6}
\end{eqnarray}

\noindent where $s'=-s$.

Equation (\ref{eq6}) is the functional relation between the real and imaginary parts of the refractive index. For the fixed value of $s'$, this relation will be single-valued. Taking into account the choice of the integer $s'$, the function $n_1=n_1(\kappa)$ of (\ref{eq6}) is multi-valued.

Solving equations (\ref{eq2}) with account of (\ref{eq4}), we have for a given incident wave

\begin{eqnarray}
E_1^0=E_0, \quad E_2^0= \frac{1}{1-\sqrt{r_1}e^{i \rho_1}} E_0, \label{eq7}
\end{eqnarray}

\begin{eqnarray}
E_3^0=- \frac{\sqrt{r_1}e^{i \rho_1}}{1-\sqrt{r_1}e^{i \rho_1}} E_0, \quad E_4^0= \sqrt{\frac{r_1}{r_2}} \frac{e^{-i \rho_2} - \sqrt{r_2}}{e^{-i \rho_1}-\sqrt{r_1}} E_0, \nonumber
\end{eqnarray}

\noindent where the incident wave amplitude $E_0$ can have any value.

The strength of electromagnetic field inside the layer is

\begin{eqnarray}
E(z,t) = \frac{E_0 e^{i \omega t}}{1-\sqrt{r_1}e^{i \rho_1}} \left[ e^{-k \kappa z - i k n_1 z} - \sqrt{r_1}e^{i \rho_1} e^{k \kappa z + i k n_1 z} \right], \label{eq8}
\end{eqnarray}

\begin{eqnarray}
H(z,t) = (n_1 - i\kappa) \frac{E_0 e^{i \omega t}}{1-\sqrt{r_1}e^{i \rho_1}} \left[ e^{-k \kappa z - i k n_1 z} + \sqrt{r_1}e^{i \rho_1} e^{k \kappa z + i k n_1 z} \right]. \nonumber
\end{eqnarray}

Additional characteristics of the electromagnetic fields can be obtained by calculating the density of electric and magnetic fields energy in the layer:

\begin{eqnarray}
W_e = \frac{|E_0|^2}{8 \pi} \frac{n_1^2 - \kappa^2}{1 + r_1 -2 \sqrt{r_1} \cos \rho_1} \left[ e^{-2k \kappa z} + r_1 e^{2k \kappa z} - 2 \sqrt{r_1} \cos(2 k n_1 z + \rho_1) \right], \label{eq9}
\end{eqnarray}

\begin{eqnarray}
W_m = \frac{|E_0|^2}{8 \pi} \frac{n_1^2 + \kappa^2}{1 + r_1 -2 \sqrt{r_1} \cos \rho_1} \left[ e^{-2k \kappa z} + r_1 e^{2k \kappa z} + 2 \sqrt{r_1} \cos(2 k n_1 z + \rho_1) \right]. \nonumber
\end{eqnarray}

\noindent The total energy density is

\begin{eqnarray}
W = \frac{|E_0|^2}{4 \pi} \frac{1}{1 + r_1 -2 \sqrt{r_1} \cos \rho_1} \times \nonumber \\ \times \left[ n_1^2 (e^{-2k \kappa z} + r_1 e^{2k \kappa z}) + 2 \kappa^2 \sqrt{r_1} \cos(2 k n_1 z + \rho_1) \right]. \label{eq10}
\end{eqnarray}

The energy flux in the layer (Poynting vector) is given by

\begin{eqnarray}
P = \frac{|E_0|^2}{4 \pi} \frac{c}{1 + r_1 -2 \sqrt{r_1} \cos \rho_1} \times \nonumber \\ \times \left[ n_1 (e^{-2k \kappa z} - r_1 e^{2k \kappa z}) + 2 \kappa \sqrt{r_1} \sin(2 k n_1 z + \rho_1) \right]. \label{eq11}
\end{eqnarray}

\noindent From (\ref{eq10}) and (\ref{eq11}), it is easy to obtain the expression

\begin{eqnarray}
\frac{dW}{dz} = -2 n_1 \kappa \frac{k}{c} P. \label{eq12}
\end{eqnarray}

\noindent It follows from (\ref{eq11}) that the derivative of energy flux over the coordinate $z$ cannot be positive, hence, the flux decreases monotonically from the first layer boundary to the second one. It is positive at the first boundary and negative at the second. There is a point inside the layer where the flux is zero. The direction of fluxes is opposite on the both sides of this point. As we move away from it, the flux magnitude increases.

Position of the point of zero flux depends on the reflection coefficients and the phase shifts at the both layer boundaries ($r_1$, $r_2$, $\rho_1$, $\rho_2$). It is generally situated not at the middle of the layer. If $r_2>r_1$ and $\rho_2>\rho_1$, the zero point is located closer to the boundary with the reflection coefficient $r_2$. In certain particular cases, it can coincide with the layer center.

It is seen from (\ref{eq12}) that the energy density is minimal at the point of zero flux and grows monotonically on the both sides of this point. The periodic terms containing trigonometric functions in (\ref{eq10}) and (\ref{eq11}) are always minor and can cause only an uneven change in the flux or energy density. Pulsations of the energy or flux are absent; they have maximal values at the layer boundaries.

Absorption inside the layer can be calculated in two ways: either with the Joule formula ($Q=\sigma E E^\ast$) or as the flux divergence taken with the opposite sign. Both formulas give the same value

\begin{eqnarray}
Q = \frac{|E_0|^2}{4 \pi} \frac{2 k c n_1 \kappa}{1 + r_1 -2 \sqrt{r_1} \cos \rho_1} \times \nonumber \\ \times \left[ e^{-2k \kappa z} + r_1 e^{2k \kappa z} - 2 \sqrt{r_1} \cos(2 k n_1 z + \rho_1) \right]. \label{eq13}
\end{eqnarray}

It follows from (\ref{eq13}) that absorption inside the layer is nonuniform and changes from point to point, but it cannot be equal to zero. The magnitude of absorption can pulsate and for low medium absorption (small $\kappa$) the number of such pulsations will be large.

In the particular case of the same limiting medium, formulas (\ref{eq10}) and (\ref{eq11}) can be significantly simplified. The functions $P$ and $W$ become symmetric with respect to the layer center. Indeed, after moving the coordinate origin to the layer center, the formulas for $P$ and $W$ can be written as

\begin{eqnarray}
W = \frac{|E_0|^2}{2 \pi} \frac{\sqrt{r_1}}{1 + r_1 -2 \sqrt{r_1} \cos \rho_1} \times \nonumber \\ \times \left[ n_1^2 \cosh 2k \kappa (z-l/2) - \kappa^2 \cos 2 k n_1 (z - l/2) \right], \nonumber
\end{eqnarray}

\begin{eqnarray}
P = -\frac{|E_0|^2}{2 \pi} \frac{\sqrt{r_1} c}{1 + r_1 -2 \sqrt{r_1} \cos \rho_1} \times \nonumber \\ \times \left[ n_1 \sinh 2k \kappa (z-l/2) + \kappa \sin 2 k n_1 (z - l/2) \right]. \label{eq14}
\end{eqnarray}

The case of resonant absorption considered here is analogous to the case of light generation in the layer with the negative absorption index (\footnote[2]{A.~P.~Khapalyuk, B.~I.~Stepanov, Vesci AN BSSR, ser. phys.-tech. (Proceedings of the National Academy of Sciences of Belarus. Physical-technical series), issue 4, p. 132, 1961.}).

In conclusion, I express my gratitude to B.~I.~Stepanov for his attention to the work and valuable advice.

\textit{Belarusian State University}

\textit{Submitted 14.III.1962}

\end{document}